\documentclass{aa}

\usepackage{graphicx}
\usepackage{txfonts}
\bibpunct{(}{)}{;}{a}{}{,} 

\usepackage{color}
\usepackage[colorlinks,linkcolor=blue,citecolor=blue,urlcolor=blue]{hyperref}
\begin{document}

\title{Equation of state SAHA-S \\
        meets stellar evolution code CESAM2k}

\titlerunning{SAHA-S meets CESAM2k}

\author{V.A. Baturin
        \inst{\ref{inst1}}
        \and    
        W. D\"appen
        \inst{\ref{inst2}}
        \and
        P. Morel
        \inst{\ref{inst3}}
        \and
        A.V. Oreshina
        \inst{\ref{inst1}}
        \and            
        F. Th\'evenin
        \inst{\ref{inst3}}
        \and                    
        V.K. Gryaznov
        \inst{\ref{inst4},\ref{inst5}}
        \and                    
        I.L.~Iosilevskiy
        \inst{\ref{inst6},\ref{inst7}}
        \and
        A.N.~Starostin          
        \inst{\ref{inst8}}
        \and
        V.E.~Fortov             
        \inst{\ref{inst4},\ref{inst6}}
}

\authorrunning{V.A. Baturin, W. D\"appen, P. Morel et al.}

\institute{Sternberg Astronomical Institute, Lomonosov Moscow State
        University, Moscow, Russia, \email{vab@sai.msu.ru}\label{inst1}
        \and
        Department of Physics and Astronomy, University of Southern
        California, Los Angeles, CA 90089, USA\label{inst2}
        \and
        Universit\'e de La Cote d'Azur, OCA, Laboratoire
        Lagrange CNRS, BP. 4229, 06304, Nice Cedex, France\label{inst3}
        \and
        Institute of Problems of Chemical Physics RAS, Chernogolovka, Russia\label{inst4}
        \and
        Tomsk State University, Tomsk, Russia\label{inst5}
        \and
        Joint Institute for High Temperatures RAS, Moscow, Russia \label{inst6}
        \and
        Moscow Institute of Physics and Technology, Dolgoprudnyi,
        Russia \label{inst7}
        \and
        Troitsk Institute for Innovation and Fusion Research, Troitsk, Russia \label{inst8}
}

   \date{Received 26 May 2017; accepted 1 August 2017}


 \abstract{}{}{}{}{}

  \abstract
   {We present an example of an interpolation code of the SAHA-S equation of state that has been adapted for use in the stellar evolution code CESAM2k.}
   {The aim is to provide the necessary data and numerical procedures for its implementation in a stellar code.  A technical problem is the discrepancy between the sets of thermodynamic quantities provided by the SAHA-S equation of state and those necessary in the CESAM2k computations. Moreover, the independent variables in a practical equation of state (like SAHA-S) are temperature and density, whereas for modelling calculations the variables temperature and pressure are preferable. Specifically for the CESAM2k code, some additional quantities and their derivatives must be provided. }
   {To provide the bridge between the equation of state and stellar modelling, we prepare auxiliary tables of the quantities that are demanded in CESAM2k. Then we use cubic spline interpolation to provide both smoothness and a good approximation of the necessary derivatives. Using the B-form of spline representation provides us with an efficient algorithm for three-dimensional interpolation.}
   {The table of B-spline coefficients provided can be directly used during stellar model calculations together with the module of cubic spline interpolation. This implementation of the SAHA-S equation of state in the CESAM2k stellar structure and evolution code has been tested on a solar model evolved to the present. A comparison with other equations of state is briefly discussed.}
   {The choice of a regular net of mesh points for specific primary quantities in the SAHA-S equation of state, together with accurate and consistently smooth tabulated values, provides an effective algorithm of interpolation in modelling calculations. The proposed module of interpolation procedures can be easily adopted in other evolution codes.}


   \maketitle
%

\section{Introduction}

One of the most important factors  for successfully modelling the stellar-interior structure is the equation of state (EOS) of the matter. It governs both the opacity and the thermodynamic structure of the star, which determine the predictions of the models, such as the surface temperature, luminosity, and radius for a given age of the star.
For several decades, several EOS formalisms have been tested in evolutionary models of different kinds of stars, mainly depending on mass
and chemical composition. For example, in the CESAM2k code \citep{Morel2008} different EOS can be considered, such as OPAL \citep{Rogers1996, RogersNayfonov2002} or the MHD EOS \citep{Mihalas1988}; the code has been
extensively used to reproduce the Sun as a star \citep{MorelEtAl1997} as well as many other stars of different mass
and chemical composition as a function of age.

In this paper we describe the implementation of the recently developed equation of state named SAHA-S \citep{Gryaznov2006, Gryaznov2013} in the CESAM2k code and provide examples of its capability to reproduce the Sun as a star. In  Sect.~\ref{SectBasicConcepts} we describe basic concepts and relations of EOS theory. The requirements by CESAM2k on the equations of state are listed in  Sect.~\ref{SectRequirementsCESAM}, and the corresponding available SAHA-S data are discussed in Sect.~\ref{SectSAHASdata}. Implementation of the SAHA-S EOS in the mathematical structure of the CESAM2k code is described in Sect.~\ref{SectBridging}. Particular attention is given to the numerical algorithms based on a three-dimensional cubic spline interpolation \citep{deBoor1978}. In Sect.~\ref{SectComparisonEOS}, we show how some physical quantities such as the sound speed profile in the interior are reproduced in a solar model based on the SAHA-S EOS, and we make comparisons with earlier CESAM2k models that were based on the OPAL EOS. Conclusions are presented in Sect.~\ref{SectConclusions}.

\section{Basic concepts and relations}
\label{SectBasicConcepts}
For matter in thermodynamical equilibrium, the equation of state is the relation between pressure $P$, temperature $T$, and density $\rho$, assuming a fixed chemical composition.
There are many introductions into
the equation of state and its role in modelling stellar structure~\citep[e.g.][]{CoxGiuli1968, Hansen1994}. However, recent computations of the equation of state based on more modern physical assumptions have therefore become very difficult and elaborate \citep[see e.g.][for the case of the SAHA-S equation of state]{Gryaznov2006, Gryaznov2013}.

\subsection{Geometry of equation of state}

Mathematically speaking, in the
coordinate space $\{P,T,\rho\}$, an equation of state can be represented as a surface (two-dimensional manifold). 
For a given mass and chemical composition, any change in the thermodynamic state is described by a curve on this surface. In models of stellar internal structure, a given trace inside the star is associated with quantities $\{P,T,\rho\}$ that lie on this EOS surface.

From a geometrical point of view, the EOS surface is a smooth (differentiable) surface. This leads to the next geometrical structure,  a tangent plane which uniquely exists at every point of the surface. To describe and define the tangent plane (which is a linear space), we write an expression for differential of pressure in the form
\begin{equation}
\mathrm{d}\ln P={{\chi }_{T}}\,\mathrm{d}\ln T+{{\chi }_{\rho }}\,\mathrm{d}\ln \rho  \, .
\end{equation}

\noindent
Following historical tradition in astrophysics, we use logarithmic variables wherever  possible. In the differential of pressure, two dimensionless partial derivatives appear
as\begin{equation}
{\left.\chi_T \equiv \frac{\partial \ln P}{\partial \ln T} \right| }_\rho
\quad \mathrm{and} \quad
{\left.\chi_\rho \equiv \frac{\partial \ln P}{\partial \ln \rho} \right| }_T \, .
\label{Eq_chiT_chiRho}
\end{equation}
These derivatives essentially describe the orientation of the tangent plane in the coordinate space. $\chi_T=\chi_\rho=1$ for a perfect-gas EOS, at least in some finite part of the EOS surface, and there the scaled pressure
\begin{equation}
\Pi\equiv P/\rho T
\label{EqScaledPressure}
\end{equation}
is constant. As a pure geometrical consequence, a third partial derivative $\delta$ can be introduced, and it is related to the two derivatives defined above:
\begin{equation}
\delta  =  - {\left. {\frac{{\partial {\mathrm{ln}}\rho }}{{\partial {\mathrm{ln}}T}}} \right|_P} = \frac{{{\chi _T}}}{{{\chi _\rho }}} \, .
\label{EqDelta}
\end{equation}

Here we provide a definition of the tangent plane with help of the pressure differential because it is the most natural for a thermodynamic description in which $T$ and $\rho$ are the independent variables. However, for applications to stellar modelling, the differential of density is more useful because $T$ and $P$ are the independent variables. Obviously, again geometrically, one can write
\begin{equation}
{\mathrm{d}}\ln \rho  =  - \delta {\mathrm{d}}\ln T + {\left( {{\chi _\rho }} \right)^{ - 1}}{\mathrm{d}}\ln P
\label{EqDRho} \, .
\end{equation}

\noindent
From this expression it becomes clear why $\delta$   and $\chi_\rho^{-1}$  are needed in stellar evolution codes, and in CESAM2k in particular.

 This picture can be generalized to the case where the coordinate space has more dimensions, as in the case of a non-constant chemical composition. This simple case  is expressed by $X$ and $Z$, which are respectively the mass fractions of hydrogen and all elements heavier than helium. With additional partial derivatives
\begin{equation}
{\left. {{\chi _X} \equiv \frac{{\partial \ln P}}{{\partial \ln X}}} \right|_{\rho ,T,Z}}\;\;\;{\kern 1pt} {\mathrm{and}}\;\;\;{\kern 1pt} {\left. {{\chi _Z} \equiv \frac{{\partial \ln P}}{{\partial \ln Z}}} \right|_{\rho ,T,X}}{\kern 1pt} ,
\label{EqChiXChiZ}
\end{equation}
the differential of density becomes
\begin{equation}
{\mathrm{d}}\ln \rho  =  - \frac{\chi_T}{\chi_\rho} {\mathrm{d}}\ln T + {\left( {{\chi _\rho }} \right)^{ - 1}}{\mathrm{d}}\ln P - \frac{{{\chi _X}}}{{{\chi _\rho }}}{\mathrm{d}}\ln X - \frac{{{\chi _Z}}}{{{\chi _\rho }}}{\mathrm{d}}\ln Z \, .
\label{EqDRhoFull}
\end{equation}
The last derivative with respect to $Z$ is smaller than the derivative with respect to $X$ and we  omit it in further expressions.

The practical meaning of the tangent plane might not be obvious until we actually deal with the principal task in the theory of the EOS. Nevertheless, let us consider  here the case where a curve $C$ in the EOS surface passes through the point $M$ characterized by the temperature-density pair ${T, \rho}$.
Then there is a tangent line to the curve $C$ at this point $M$. This tangent line is also part of the tangent plane of the EOS surface originating at the point $M$.
 If we know one coordinate projection of the tangent line, this means that we can deduce the two other projections as well.

In the next chapter, we  consider an important illustration of this problem (the adiabatic process).

\subsection{Adiabatic gradients}

The  fundamental thermodynamic laws state that there is a unique curve of an isentropic process for each point of the EOS, and again the direction of adiabatic changes belongs to the tangent plane of that point. Generally,
 adiabatic changes performed under constant chemical composition are considered. The adiabatic direction can be determined by any of the three projections on the coordinate planes. Conventionally, the best-known notations for them are
\begin{equation}
{\Gamma _1} \equiv {\left. {\frac{{\partial \ln P}}{{\partial \ln \rho }}} \right|_S},
\end{equation}
\begin{equation}
{\nabla _{{\mathrm{ad}}}} \equiv {\left. {\frac{{\partial \ln T}}{{\partial \ln P}}} \right|_S},
\end{equation}
\begin{equation}
{\Gamma _3} - 1 \equiv {\left. {\frac{{\partial \ln T}}{{\partial \ln \rho }}} \right|_S} \, ,
\end{equation}

\noindent
where $\nabla_{\mathrm{ad}}$  is the standard notation for the adiabatic temperature gradient, while the notations $\Gamma_1$   and $\Gamma_3-1$  refer to S. Chandrasekhar  (\cite{Chandrasekhar1939}, \cite{Lang1974} or \cite{CoxGiuli1968}, \cite{Weiss2004}).
The three derivatives are obviously related

\begin{equation}
{\Gamma _3} - 1= {\Gamma _1}{\nabla _{{\mathrm{ad}}}} \, .
\end{equation}
However, to obtain any of these adiabatic projections  additional thermodynamic information is required. The most natural way is based on the caloric equation of state, expressed through
\begin{equation}
{\Gamma _3} - 1 = \Pi \frac{{{\chi _T}}}{{{c_V}}}
\label{EqGamma3}
,\end{equation}
where $\Pi$ is defined by Eq.~(\ref{EqScaledPressure}), $c_V$ is the specific heat capacity at constant volume
\begin{equation}
{c_V} = T{\left. {\frac{{\partial S}}{{\partial T}}} \right|_V} = {\left. {\frac{{\partial U}}{{\partial T}}} \right|_V}{\kern 1pt} .
\end{equation}
If the adiabatic projection $\Gamma_3-1$  is found, the other two  become
\begin{equation}
\nabla _{\mathrm{ad}}^{ - 1} = \frac{{{\chi _\rho }}}{{{\Gamma _3} - 1}} + {\chi _T}
\end{equation}
and
\begin{equation}
{\Gamma _1} = {\chi _T}\left( {{\Gamma _3} - 1} \right) + {\chi _\rho }\, .
\end{equation}
The value of $\nabla_{\mathrm{ad}}$  is fundamental for modelling stellar convection zones, whereas $\Gamma_1$ is directly related to the adiabatic sound speed
${c^2} = {{{\Gamma _1}P} \mathord{\left/ {\vphantom {{{\Gamma _1}P} \rho }} \right. \kern-\nulldelimiterspace} \rho }$ ,
which is necessary for seismic calculations.
Also, there is a direct relation between $\Gamma_1$  and $\nabla_{\mathrm{ad}}$ without recourse to $\Gamma_3-1$:
\begin{equation}
\Gamma _1^{ - 1} = \frac{{1 - {\chi _T}{\nabla _{\mathrm{ad}}}}}{{{\chi _\rho }}} \, .
\label{EqGamma1viaNablaAd}
\end{equation}
In addition to $c_V$, another caloric quantity is $c_P$, the specific heat capacity under constant pressure:
\begin{equation}
{c_P} = T{\left. {\frac{{\partial S}}{{\partial T}}} \right|_P}{\kern 1pt} \, .
\end{equation}
A common way to calculate $c_P$ is by using the difference between the two capacities, which is solely a consequence of the geometry of the EOS surface:
\begin{equation}
{c_P} - {c_V} = \Pi \frac{{\chi _T^2}}{{{\chi _\rho }}}{\kern 1pt} \, .
\label{EqCp}
\end{equation}
Then the adiabatic exponents are
\begin{equation}
{\nabla _{\mathrm{ad}}} = \frac{\Pi }{{{c_P}}}\frac{\chi _T}{\chi_\rho} \,
\label{EqNablaAd}
\end{equation}
and
\begin{equation}
\Gamma_1=\frac{c_P}{c_V}\chi_{\rho} \, .
\label{EqGamma1}
\end{equation}
In summary, the minimum necessary thermodynamic information is given by the three quantities $P$, $\chi_T$, $\chi_{\rho }$, and by one of four quantities
$\left( {{c_V},{c_P},{\nabla _{\mathrm{ad}}},{\Gamma _1}} \right)$.
Specifically, the stellar evolution code CESAM2k requires the data set $\rho ,U,\delta ,{c_P},{\nabla _{\mathrm{ad}}},\chi _\rho ^{ - 1},{\Gamma _1}$, which can be obtained from that set of thermodynamic quantities.

\subsection{Trace of the stellar model on the EOS surface}

To avoid going into too much detail about the modelling procedure, here we outline only the general scheme. The modelling is principally based on equations giving $\nabla P$  and $\nabla T$  as functions of $T,\rho ,X,Z$ at every point. The goal of the procedure is to obtain a model curve $C_{\mathrm{M}}$, expressed as the parametric functions $P\left( r \right),T\left( r \right),\rho \left( r \right)$, where $r$ is radial coordinate in the model. This curve belongs to the general EOS surface in $\left\{ {PT\rho ;XZ} \right\}$  space.
The derivative along the radius of the model curve $C_{\mathrm{M}}$
\begin{equation}
{\nabla _{\mathrm{M}}} = \frac{{d\ln T}}{{d\ln P}}
\label{EqNablaM}
\end{equation}
depends not only  on local thermodynamic values, but also on model parameters. Therefore, the individual points on the model curve can only be determined as a result of the calculation of the entire model.

The spatial derivative of the density $\mathrm{d}\rho /\mathrm{d}r$ is necessary for the calculation of the square of the Brunt-V\"ais\"al\"a frequency $N_{\mathrm{BV}}^2$, which together with the square of the sound speed
$c_s^2$,
\begin{equation}
c_S^2={\left. {\frac{\partial P}{\partial \rho}} \right|}_S =
\frac{P}{\rho}\, \Gamma_1 \,
,\end{equation}
constitutes the principal parameters defining the oscillation properties of the model. By definition \citep{Unno1989},
\begin{equation}
\frac{{N_{{\mathrm{BV}}}^2}}{g} = \left( {\frac{{{\mathrm{d}}\ln P}}{{{\mathrm{d}}r}}\Gamma _1^{ - 1} - \frac{{{\mathrm{d}}\ln \rho }}{{{\mathrm{d}}r}}} \right)\, ,
\label{EqBVdefinition}
\end{equation}
where $g$  is the gravitational acceleration. Using spatial derivatives in the model $\widetilde \Gamma _{\mathrm{M}}$
\begin{equation}
{\widetilde \Gamma _{\mathrm{M}}} = \frac{\mathrm{d}\ln P}{\mathrm{d}\ln \rho}
\end{equation}
we can write
\begin{equation}
\frac{{N_{\mathrm{BV}}^2}}{g} = \frac{{{\mathrm{d}}\ln P}}{{{\mathrm{d}}r}}\left( {\Gamma _1^{ - 1} - \widetilde \Gamma _{\mathrm{M}}^{ - 1}} \right) \, .
\label{EqBVfrequency}
\end{equation}

The structure of $\widetilde \Gamma _{\mathrm{M}}$ is similar to the structure of $\nabla_{\mathrm{M}}$ (Eq.~(\ref{EqNablaM})), except that in the case of $\widetilde \Gamma _{\mathrm{M}}$ we also need the gradient of density $\nabla\rho$ in the model,  but $\nabla \rho$ is not directly available, in contrast to $\nabla P$ and $\nabla T$ which are the part of the modelling procedure.
There are two ways to get $\nabla\rho$. One is by direct numerical differentiation of model profile $\rho(r)$, but that may be numerically unstable. Another way is by expression of the thermodynamical differential for density (\ref{EqDRhoFull}). Then an expression for $\widetilde \Gamma _{\mathrm{M}}$  can be obtained in the form
\begin{equation}
\widetilde \Gamma _{\mathrm{M}}^{ - 1} =  - \delta {\nabla _{\mathrm{M}}} + {\left( {{\chi _\rho }} \right)^{ - 1}} - \frac{{{\chi _X}}}{{{\chi _\rho }}}\frac{{{\mathrm{d}}\ln X}}{{{\mathrm{d}}\ln P}} \, .
\label{EqGammaM}
\end{equation}
In deriving Eq.~(\ref{EqGammaM}) from Eq.~(\ref{EqDRhoFull}), we have omitted the last term in (\ref{EqDRhoFull}), containing the derivative of pressure with respect to $Z$, i.e. $\chi_Z$.
Nevertheless, we still use and need the derivative of pressure $\chi_X$, together with the gradient of $X(r)$ from the model data.

Generally, the gradient of $X$  is a much more slowly varying function than the gradient of $\rho$. In several important regions of the star it is equal to or close to zero. For example, in the mixed and chemically homogeneous convection zone, the following expression for $N_{\mathrm{BV}}^2$ can be used
\begin{equation}
\frac{{N_{\mathrm{BV}}^2}}{g} = \frac{{g{\Gamma _1}}}{{{c_S^2}}}\frac{{{\chi _T}}}{{{\chi _\rho }}}\left( {{\nabla _{\mathrm{ad}}} - {\nabla _{\mathrm{M}}}} \right)
\label{EqBVcz}
,\end{equation}
where Eqs.~(\ref{EqGamma1viaNablaAd}) and (\ref{EqDRho}) have been taken into account. In the convection zone, expression~(\ref{EqBVcz}) is much more stable and virtually exact when compared with the same quantity obtained by numerical differentiation. The reason is that the difference $\left( {{\nabla _{\mathrm{ad}}} - {\nabla _{\mathrm{M}}}} \right)$ is very small in most places in the convection zone,  and in addition is directly provided by convective transport equation.
However, Eq.~(\ref{EqGammaM}) is so far not used in CESAM2k, and Eq.~(\ref{EqBVdefinition}) is used in the convection zone and in other parts of the star.

\subsection{Calculation of gravitational energy contribution}

In calculations of stellar evolutions, the contribution from the release of gravitational energy is significant, especially on the pre-main sequence stage. It is calculated on the basis of the formula
\begin{equation}
\varepsilon_{\mathrm{grav}}=T\frac{\mathrm{d}S}{\mathrm{d}t} \, ,
\end{equation}
\noindent where
\begin{equation}
T\mathrm{d}S=\mathrm{d}U+P\mathrm{d}V \, .
\end{equation}

\noindent The computation of $\mathrm{d}U$ requires the internal energy $U$ from the equation of state.

\section{Requirements of CESAM2k regarding the thermodynamic quantities of the equation of state}
\label{SectRequirementsCESAM}

The stellar evolution code CESAM2k \citep{Morel2008} calls for an equation of state in the form of
$ \rho = \rho (P, T)$,
i.e. with  pressure $P$ and temperature $T$ as the independent variables. Further input parameters for the physical quantities are the hydrogen mass fraction $X$ and the mass fraction $Z$ of all elements heavier than helium. Additionally, there are auxiliary thermodynamic derivatives specifically requested in CESAM2k calculations. These additional derivatives are listed in Table~1 and marked there with asterisks. Historically, for the sake of time saving, the logical parameter {\tt deriv} was used to control the volume of these calculations. During the present CESAM2k calculations, the EOS procedure can still be performed with {\tt deriv=.true.} or {\tt .false.} should certain output quantities be desired. 
Output parameters of the equation of state in CESAM are the density $\rho$, together with a wide set of thermodynamic quantities:  internal energy $U$, specific heat $c_P$, adiabatic gradient $\nabla_{\mathrm{ad}}$, adiabatic exponent $\Gamma_1$, and also the values of $\delta$, $\alpha$, $\beta$:
\begin{equation}
\delta\equiv\chi_T/\chi_\rho
\end{equation}
\begin{equation}
\alpha
\equiv 1/\chi_\rho \, ,
\label{EqAlfa}
\end{equation}
\begin{equation}
\beta \equiv 1 - (a/3)T^4/P \, ,
\label{EqBeta}
\end{equation}
\noindent where $a$ is the radiation constant ($a\approx 7.5657\cdot10^{-15}$ erg cm$^{-3}$ K$^{-4}$). The derivatives of these functions with respect to $T$ at constant $P$, and to $P$ at constant $T$ are also required, together with the derivatives with respect to the hydrogen abundance $X$. Details of the interface are listed in Table~\ref{TableInterface}.

\begin{table}
        \caption{Interface of CESAM2k and an equation of state}
        \label{TableInterface}
        \begin{tabular}{ll}
                \hline
                Input:  & \\
                Pressure $P$ &  \\
                Temperature $T$ &  \\
                \multicolumn{2}{l}{Hydrogen mass fraction $X$}  \\
                \multicolumn{2}{l}{Heavy-element mass fraction $Z$}  \\
            \multicolumn{2}{l}{Logical parameter {\tt deriv} for derivatives\tablefootmark{*}} \\
                \hline
                Output: & \\
                Variable & Computation in SAHA-S module \\
                \hline
                $\rho$  & Inverse interpolation $P(T,Q_s,X)$ \\
                $\partial\rho /\partial P|_{T,X}$  &  Eq.~(\ref{EqDrop})  \\
                $\partial\rho /\partial T|_{P,X}$  &  Eq.~(\ref{EqDrot})  \\
                $\partial\rho /\partial X|_{P,T}$  &  Eq.~(\ref{EqDrox})  \\
                $U$           & Originally available SAHA-S data \\
                $\partial U /\partial P|_{T,X}$  &  Eq.~(\ref{EqDup})  \\
                $\partial U /\partial T|_{P,X}$  & Eq.~(\ref{EqDut})   \\
                $\partial U /\partial X|_{P,T}$  &  Auxiliary SAHA-S data  \\
                $ \delta $  & Auxiliary SAHA-S data \\
                $\partial \delta /\partial P|_{T,X}$\tablefootmark{*}  & Spline derivatives, Eq.~(\ref{EqDfP})   \\
                $\partial \delta /\partial T|_{P,X}$\tablefootmark{*}  & Spline derivatives, Eq.~ (\ref{EqDfT}) \\
                $\partial \delta /\partial X|_{P,T}$\tablefootmark{*}  &  Spline derivative,  Eq.~(\ref{EqDfx})  \\
                $c_P$                       & Auxiliary  SAHA-S data \\
                $\partial c_P /\partial P|_{T,X}$\tablefootmark{*} & Spline derivatives, Eq.~(\ref{EqDfP})\\
                $\partial c_P /\partial T|_{P,X}$\tablefootmark{*} & Spline derivatives, Eq.~(\ref{EqDfT}) \\
                $\partial c_P /\partial X|_{P,T}$\tablefootmark{*} & Spline derivative, Eq.~(\ref{EqDfx}) \\
                $\nabla_{\mathrm{ad}}$ & Auxiliary SAHA-S data \\
                $\partial \nabla_\mathrm{ad} /\partial P|_{T,X}$\tablefootmark{*} & Spline derivatives, Eq.~(\ref{EqDfP}) \\
                $\partial \nabla_\mathrm{ad} /\partial T|_{P,X}$\tablefootmark{*} & Spline derivatives, Eq.~(\ref{EqDfT}) \\
                $\partial \nabla_\mathrm{ad} /\partial X|_{P,T}$\tablefootmark{*} & Spline derivative, Eq.~(\ref{EqDfx}) \\
                $\alpha$ &  Eq.~(\ref{EqAlfa})\\
                $\beta$ &  Eq.~(\ref{EqBeta})\\
                $\Gamma_1$ & Auxiliary SAHA-S data\\
                \hline
        \end{tabular}
\tablefoot{Derivatives marked by asterisk are computed only if {\tt deriv=.TRUE.}}
\end{table}


\section{Available SAHA-S data}
\label{SectSAHASdata}

The physical description of the SAHA-S equation of state is found in \cite{Gryaznov2006, Gryaznov2013}. The currently available data are described in \cite{Baturin2013}. They are in
original form, by which we mean that their physical input parameters are based on temperature $T$ and density $\rho$. This means that  they describe the equation of state in the form $P=P(T, \rho)$, which is different from the form of the equation of state as required by CESAM2k.

More specifically, the available SAHA-S data are tabulated with the parameters $T$ and $Q_s(\rho , T)$:
\begin{equation}
Q_s = \frac{\rho}{(T/10^6)^{2.25}} \, .
\label{Eq_Qs}
\end{equation}

\begin{figure}
        \centering
        \resizebox{\hsize}{!}{\includegraphics{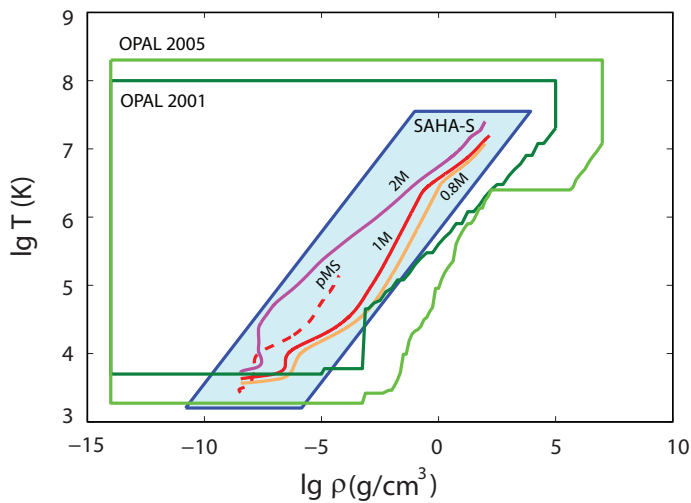}}
        \caption{Domain of applicability of the SAHA-S EOS, and two versions of the OPAL EOS. The profiles of main sequence stars with masses $0.8M_{\odot}$, $1.0M_{\odot}$, and $2.0M_{\odot}$ are indicated by the solid orange, red, and magenta curves, respectively. The dashed curve shows the profile of a $1M_{\odot}$ early pre-main sequence model.}
        \label{FigDomains}%
\end{figure}

\noindent The  power 2.25 was used in Eq.~(\ref{Eq_Qs}) because, during the creation of the SAHA-S basic data tables, it leads to the most compact rectangular coverage of the solar model trace (see Fig.~\ref{FigDomains}). At the same time, with that choice, the authors managed to avoid the problematic region of extremely high density which is not relevant to ordinary stars. We note that in future versions of SAHA-S this choice of power might be changed to allow modelling a broader mass range of stars. 

The (bijective) transformation above allows us to calculate stellar models with the help of
{rectangular} tables, which is very useful for the necessary interpolations within the tables (see Sect.~\ref{SectInterpolation}).
In particular, the choice of a new independent variable $Q_s$ instead of $\rho$ makes it possible to interpolate in a two-dimensional, equidistant, and rectangular grid of mesh points. This has the advantage that the two-dimensional mesh can be written as a product of two one-dimensional meshes. In the plane of $T-\rho$, the SAHA-S mesh points lie inside a parallelogram limited by the blue boundaries in Fig.~\ref{FigDomains}.
Compared with the boundaries of the OPAL tables (also plotted in Fig.~\ref{FigDomains}), those of SAHA-S are different because it was specifically developed for solar modeling; however, the table construction of SAHA-S is also necessary for our multi-dimensional non-local spline interpolation, which makes this procedure effective and simple.
The track of solar model points is shown by the solid red curve. Solar evolution models can be calculated onward from an early stage of the pre-main sequence, an example of such a calculation is plotted by the dashed red curve.

To assess the possibility of using the SAHA-S tables in its presently available domain to stars of different mass, we have plotted two examples, one for a star with $0.8M_\odot$ (orange curve) and the other for a star with $2M_\odot$, close to the end of their main sequence lives (magenta curve).

As shown in Fig.~\ref{FigDomains}, we conclude that for low-mass stars, modelling with the SAHA-S EOS is rather limited, whereas for massive main sequence stars it has better prospects. However, for such applications the domain of the tables will have to be expanded toward higher temperatures and lower densities.

As mentioned above, two additional input parameters of SAHA-S tables are the hydrogen mass fraction $X$ and the heavy-element mass fraction $Z$. With respect to $X$, we  perform a B-spline interpolation, but in $Z$ our interpolation is linear.
We note that the mixture of the heavy elements included in SAHA-S cannot be changed unless  a new set of thermodynamic tables is  constructed. The current version of SAHA-S EOS includes the eight most abundant elements heavier than helium.

The original SAHA-S thermodynamic quantities  provide all the necessary thermodynamic information, but not in the form of the quantities needed by CESAM2k. While SAHA-S provides $P$, $\chi_T$, $\chi_\rho$,  $c_V$, together with the derivative with respect to the chemical composition $(\partial P / \partial X)|_{T,\rho}$, and also the quantities $U$, $(\partial U / \partial \rho)|_{T,X}$, and $(\partial U / \partial X)|_{T, \rho}$, CESAM2k requirements include additional quantities that have to be calculated, such as $\delta$, $c_P$, $\nabla_{{\mathrm{ad}}}$, and their derivatives.

\section{Bridging between SAHA-S and CESAM2k}
\label{SectBridging}

To integrate SAHA-S into CESAM2k, we had to apply several modifications. First, a transformation from $P(T,\rho)$  to $\rho (P,T)$ has been performed (see Sect.~\ref{SectTransformRhoPT}). Second, the set of original SAHA-S thermodynamic quantities has been expanded to match the requirements from CESAM2k. The additional quantities are listed in Table~\ref{TableAuxiliarySAHAS}.

\begin{table}
        \caption{Auxiliary data of SAHA-S}
        \label{TableAuxiliarySAHAS}
        \begin{tabular}{ll}
                \hline
                 Variable & Computation  \\
        \hline
           $P$, dyn/cm$^2$  \hphantom{0000}  & Original SAHA-S data \\
           $U$, erg/g                        & Original SAHA-S data \\
       $\partial U/\partial\rho \,|_{T}$ & Original SAHA-S data \\
       $c_P$                             & Eq.~(\ref{EqCp})\\
       $\chi_\rho$                       & Original SAHA-S data\\
       $\delta $                         & Eq.~(\ref{EqDelta})\\
           $\Gamma_1$                        & Eq.~(\ref{EqGamma1}) \\
           $\nabla_{\mathrm{ad}}$            & Eq.~(\ref{EqNablaAd})\\
           $\partial U/\partial X \, |_{P,\,T}$ & Eq.~(\ref{EqDux})\\
           $\rho$, g/cm$^3$                 & Original SAHA-S data\\
           $\partial \ln P/\partial \ln X \, |_{T,\,\rho}$ & Original SAHA-S data\\
           \hline
\end{tabular}
\end{table}

Third, a B-spline interpolation of the resulting expanded tables has been performed in the variables $\log(T)$, $\log(Q_s)$, and $X$.  The advantage of this type of interpolation is its high efficiency even in three dimensions.  Also, it allows the calculation of the derivatives of all tabulated functions and of functions based on the tabulated quantities. Moreover, a B-spline interpolation preserves the smoothness of the thermodynamic functions. Details of the technique are discussed in Sect.~\ref{SectInterpolation}.

Fourth, and finally, we have developed auxiliary routines to provide the transformation from derivatives with respect to $Q_s$ to derivatives with respect to $\rho$, and from derivatives with respect to $T$, $\rho$ to derivatives with respect to $P$, $T$.
Our routines use B-spline derivatives with respect to $X$ under constant $T$ and $Q_s$, which are transformed to those at constant $P$ and $T$ (see Sect.~\ref{SectDerivatives}). Previously, in CESAM2k only a finite difference differentiation was used for the latter, requiring extra calls of the EOS procedure.

\subsection{Transformation of the equation of state to the form $\rho(P,T)$}
\label{SectTransformRhoPT}

When starting from $P = P\left( {\rho ,T} \right)$, represented in the form of a set of piece-wise interpolation polynomials of type ${}^{\left( j \right)}S\left( x \right)$ (with $j$ being the order),  a method to find the inverse functions of these interpolation polynomials
${}^{\left( j \right)}{S^{ - 1}}\left( y \right)$ is needed.
In the linear case ($j=1$), the solution of this problem is simple and analytical, but in the case of quadratic and cubic polynomials it results in complicated irrational functions. For cubic splines, an analytical inversion  becomes practically useless and numerically inefficient. We
therefore solve the inversion problem numerically by the Newton-Raphson iteration method \citep{Press1992}. In particular, for a given $P_0$ we must find  the root $\rho_0$ of the equation
$P_0 = P\left( \rho  \right)$. The value of the density in step $i$ of the iteration is
\begin{equation}
\rho ^{(i)} = \rho ^{(i-1)} +  \frac{1}{P^\prime (\rho)}
\left[ P_0 - P(\rho ^{(i - 1)})\right]\, .
\end{equation}

\noindent The efficiency of the iterative method strongly depends on the quality of the evaluation of the derivative
$P^\prime (\rho)$ during every step of the iteration. In addition to the rather crude but robust method based on finite-difference estimates of the derivative, two other approaches for $P^\prime (\rho)$ are available. The first uses the derivative $ \left.\partial P/ \partial \rho \,\right|_T $
obtained from the tabulated data of equation of state. This method achieves a rapid convergence of the iteration, but it requires an additional interpolation of the quantity
${\chi _\rho }\left( {T,\rho ,X} \right)$. The second mathematically more rigorous method consists in obtaining the derivative from the analytical  cubic polynomial for
${P\left( \rho  \right)}$ with given fixed $T,X$. In the present paper, we  use the method based on interpolation of $\chi_{\rho }$.

The starting point of the density iteration $\rho ^{(0)}$ is a bit tricky, and its choice can affect the convergence rate and number of iterations necessary.
Using the inverse linear interpolation function between a pair of mesh points lying nearest to the desired pressure value $P = {\left. {P\left( {{\rho _j}} \right)} \right|_{T,X}}$  for a given value of $T$and $X$, we obtain a reasonably good starting point of the iteration
\begin{equation}
\rho ^{(0)} = t{\rho _j} + (1 - t){\rho _{j + 1}}\, ,
\end{equation}

\noindent where
$t =    \left[ P - P(\rho _{j + 1}) \right] \left/ \,
        \left[ P(\rho_j) - P(\rho_{j + 1}) \right]\right.$
in the interval around $P$,     
$P(\rho _j) < P < P(\rho _{j + 1})$.


\subsection{Interpolation technique}
\label{SectInterpolation}

To represent thermodynamic  functions, we use the cubic spline interpolation with special end conditions, known as the \emph{not-a-knot condition}, following \cite{deBoor1975}.
Using the \cite{deBoor1978} notation,  a cubic spline will have order $k=4$.

We interpolate the function $f(x_i)$, given on the equidistant mesh $x_i$, $i = 1...N$, in the interval $[a; b]$.
We use a B-form of the presentation of the splines, where the result is a sum over $N$ basis functions $B_{j,k}(x)$:
\begin{equation}
S(x) = \sum\limits_{j=1}^{N} {\alpha _j}B_{j,k}(x) \, .
\end{equation}

\noindent In this representation $\alpha_j$ are dependent on $f(x)$, but $B_{j,k}(x)$ are not. So, to interpolate a number of functions at the same point we need to calculate $B_{j,k}(x)$
only once. Moreover B-forms can be easily expanded to multi-dimensional interpolation:

\begin{equation}
S(x,y,z) = \sum\limits_{j1,j2,j3} \alpha_{j1,j2,j3} B_{j1,k}(x)B_{j2,k}(y)B_{j3,k}(z) \, .
\end{equation}

We proceed to calculate basis functions, $B_{j,k}(x)$, which are used to interpolate to point $x$.

The definition of $B_{j,k}(x)$ depends on a mesh of spline-sites $t_j$ which is defined not uniquely. Generally, $t_j$ may not coincide with any knots of interpolation mesh  $x_j$.
For the sake of simplicity, we construct $t$-mesh as follows: 

\begin{equation}
\begin{array}{lcl}
t_1=x_1, & & t_{N+4}=x_N, \\
t_2=x_1, & & t_{N+3}=x_N, \\
t_3=x_1, & & t_{N+2}=x_N, \\
t_4=x_1, & t_5=x_3, \quad t_6=x_4 \quad ...\quad t_N=x_{N-2}, & t_{N+1}=x_N.  \\
\end{array}
\end{equation}

\begin{figure}
        \centering
        \resizebox{\hsize}{!}{\includegraphics{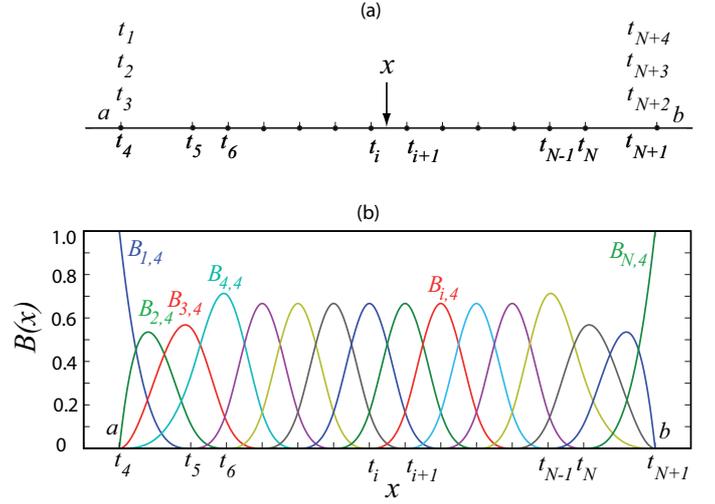}}
        \caption{(a) Mesh $\{t_i\}$ with not-a-knot end condition, (b) cubic B-spline basis functions $B_{i,k}(x)$.}
        \label{FigBspline}
\end{figure}

\noindent For not-a-knot condition, the first four and the last four knots have multiplicity 4, and the second ($i=2$) and penultimate ($i=N-1$) $x$-knots are omitted (Fig.~\ref{FigBspline}(a)). The total number of $t$-mesh points is $N+k$.

For a given $x$, let us define integer  {\it index of interval} $i(x)$ in $t$-mesh  according to
two conditions: $t_i< t_{i+1}$ and $x \in [t_i, t_{i+1})$. The first condition in definition of $i$ is important in the case of multiplicative knots.
To compute the interpolating polynomial $S(x)$ at the point $x$ within the
interval $i$, only $k$ non-zero basis functions (and corresponding coefficients $\alpha_i$) are needed:
\begin{equation}
S(x) = \sum\limits_{j=i-k+1}^{i} {\alpha _j}B_{j,4}(x) \, .
\end{equation}

To calculate all needed basis functions at $x$, it is most efficient to use the recursion relation \citep{deBoor1978}:
\begin{equation}
B_{i,k}(x) = \frac{x-t_i}{t_{i+k-1}-t_i}B_{i,k-1}(x) +
\frac{t_{i+k}-x}{t_{i+k}-t_{i+1}}B_{i+1,k-1}(x) \, .
\label{EqBasicFunctions}
\end{equation}

\noindent The recursion is started from the first-order spline
\begin{equation}
B_{i,1}(x)= \left\{
\begin{array}{l}
1 \quad \mathrm{ if } \quad  t_i\le x < t_{i+1} \, , \\
0 \quad \mathrm{elsewhere.}
\end{array}     
\right.
\end{equation}

\noindent  Sometimes Eq.~(\ref{EqBasicFunctions}) is considered the definition of the basis $B$-functions.

Figure~\ref{FigBspline} shows the $t$-mesh and the B-functions calculated for our case interpolation procedure described above. We note that the B-functions depend on the specific choice of the $t$-knots. 

Another property of B-splines is the possibility of calculating analytical derivatives of the
interpolating polynomials in the form of giving lower order splines.
For example, the first derivative ($j=1$ in Eq.~12b \cite{deBoor1978}) can be written in the form
\begin{equation}{D^1}\left( {\sum\limits_i {{\alpha _i}{B_{i,k}}} } \right) = \sum\limits_i {\alpha _i^{\left( 2 \right)}{B_{i,k-1}}} \, , \end{equation}

\noindent where the quadratic splines $B_{i,k-1}$ have already been calculated and used in Eq.~(\ref{EqBasicFunctions}), and where
\begin{equation}
\alpha _i^{(2)} =
        \frac{\alpha_i - \alpha_{i - 1}}{{(t_{i+k-1} - t_i)} /(k-1)} .
\end{equation}

\noindent For all the functions listed in Table~\ref{TableAuxiliarySAHAS}, files with the appropriate B-spline coefficients for interpolation and differentiation in $\log T$, $\log Q_s$, and $X$, are included in the SAHA-S module.

\subsection{Algorithm for calculation of thermodynamic derivatives}
\label{SectDerivatives}

To incorporate the SAHA-S equation of state into a stellar evolution code, the thermodynamic derivatives with respect to $P$, $T$, and $X$, are needed, as listed in Table~\ref{TableInterface}. The first group is the derivatives of density, and they can be extracted from the terms of Eq.~(\ref{EqDRhoFull}): 

\begin{equation}
{\left. {\frac{{\partial \rho }}{{\partial P}}} \right|_T}  = \frac{\rho }{P}\,\chi _\rho ^{ - 1}\, ,
\label{EqDrop}
 \end{equation}

\begin{equation}{\left. {\frac{{\partial \rho }}{{\partial T}}} \right|_P} =   - \frac{\rho }{T}\delta \, ,
\label{EqDrot}
\end{equation}

\begin{equation}
{\left. {\frac{{\partial \rho }}{{\partial X}}} \right|_{P,T}} =  -\frac{\rho }{X}\, \chi _\rho ^{- 1} \chi_X\, .
\label{EqDrox}
\end{equation}

\noindent In all these derivatives of density, we have used $\chi_\rho^{-1}$  and $\delta$ instead of the basic set of derivatives in Eq.~(\ref{EqDRhoFull}). The derivative
${\left. {{{\partial \rho }}/{{\partial X}}} \right|_{P,T}}$ in Eq.~(\ref{EqDrox}) is expressed via $\chi_X$ defined by Eq.~(\ref{EqChiXChiZ}).

Second group of derivatives is of internal energy $U$:

\begin{equation}
{\left. {\frac{{\partial U}}{{\partial T}}} \right|_P} = T{\left. {\frac{{\partial S}}{{\partial T}}} \right|_P} - P{\left. {\frac{{\partial V}}{{\partial T}}} \right|_P} = {c_P} - \Pi \delta \, .
\label{EqDut}
\end{equation}  

\noindent
Other symmetrical derivatives can be found in an analogous way:

\begin{equation}
        {\left. {\frac{{\partial U}}{{\partial P}}} \right|_T} = T{\left. {\frac{{\partial S}}{{\partial P}}} \right|_T} - P{\left. {\frac{{\partial V}}{{\partial P}}} \right|_T} =   \frac{1}{\rho }\left( { - \delta  + \chi _\rho ^{ - 1}} \right).
\label{EqDup}
\end{equation}
So the corresponding procedure can be simplified and uses only tabulated values $c_P$, $\chi_\rho$, and $\delta$.
The last derivative of $U$  with respect to $X$ is expressed via values originally available in SAHA-S table:
\begin{equation}
{\left. {\frac{{\partial U}}{{\partial X}}} \right|_{P,T}} = {\left. {\frac{{\partial U}}{{\partial X}}} \right|_{\rho ,T}} + {\left. {\frac{{\partial U}}{{\partial \rho }}} \right|_{X,T}}{\left. {\frac{{\partial \rho }}{{\partial X}}} \right|_{P,T}}.
\label{EqDux}
\end{equation}

To transform the derivatives from the SAHA-S coordinates of $\log Q_s$, $\log T$, and $X$ to the CESAM2k coordinates of $P$, $T$, and $X$, we use

\begin{equation}
{\left. {\frac{{\partial f}}{{\partial X}}} \right|_{P,T}} = {\left. {\frac{{\partial f}}{{\partial X}}} \right|_{\rho ,T}} + {\left. {\frac{{\partial f}}{{\partial \ln Q_s}}} \right|_{X,T}}\,\,\frac{1}{{\rho}}\,\,{\left. {\frac{{\partial \rho }}{{\partial X}}} \right|_{P,T}} \, ,
\label{EqDfx}
\end{equation}

\begin{equation}
\left.{\frac{\partial f}{\partial T}} \right|_{P,X} =
\left(
\left.{\frac{\partial f}{\partial \ln T }}\right|_{Q_s,X}  -
(\delta+2.25) \left.{\frac{\partial f}{\partial \ln Q_s}}\right|_{T,X}
\right)\frac{1}{T} \, ,
\label{EqDfT}
\end{equation}

\begin{equation}
{\left. {\frac{{\partial f}}{{\partial P}}} \right|_{T,X}} = {\left. {\frac{{\partial f}}{{\partial \ln Q_s}}} \right|_{T,X}} \,\,
{\left. {\frac{{\partial \rho }}{{\partial P}}} \right|_{T,X}} \,\,
\frac{1}{\rho} =
{\left. {\frac{{\partial f}}{{\partial \ln Q_s}}} \right|_{T,X}}\,\,\frac{1}{{P}}\,\,\chi _\rho ^{ - 1} \, ,
\label{EqDfP}
\end{equation}  

\noindent where $f$ is any of the quantities from Table~\ref{TableAuxiliarySAHAS}.

The expression of derivatives with respect to  $\ln Q_s ,\ln T$ through derivatives with respect to $\ln\rho,\ln T$ is given by

\begin{equation}
{\left. {\frac{{\partial f}}{{\partial \rho }}} \right|_T} = {\left. {\frac{{\partial f}}{{\partial \ln Q_s}}} \right|_T}\,\, \frac{1}{{\rho}} \, ,
\end{equation}

\begin{equation}{\left. {\frac{{\partial f}}{{\partial \ln T}}} \right|_\rho } = {\left. {\frac{{\partial f}}{{\partial \ln T}}} \right|_{Q_s}} + {\left. {\frac{{\partial f}}{{\partial \ln Q_s}}} \right|_T}\left( { - 2.25} \right) \, .
\end{equation}

\section{Comparison of the SAHA-S and OPAL 2001 equations of state incorporated in CESAM2k}
\label{SectComparisonEOS}

To demonstrate the correct working of SAHA-S inside CESAM2k, we present the results of computations performed with the stellar evolution code CESAM2k. All figures illustrate the profiles of physical quantities in calibrated models of the Sun. These models differ in the equation of state used: OPAL 2001 and SAHA-S, respectively.

The SAHA-S equation of state has been incorporated in CESAM2k with the aforementioned interpolation method, which is numerically efficient and, in addition, inside the model it allows interpolation in $Z$ at every point in space and time.

\begin{figure}
        \centering
        \resizebox{\hsize}{!}{\includegraphics{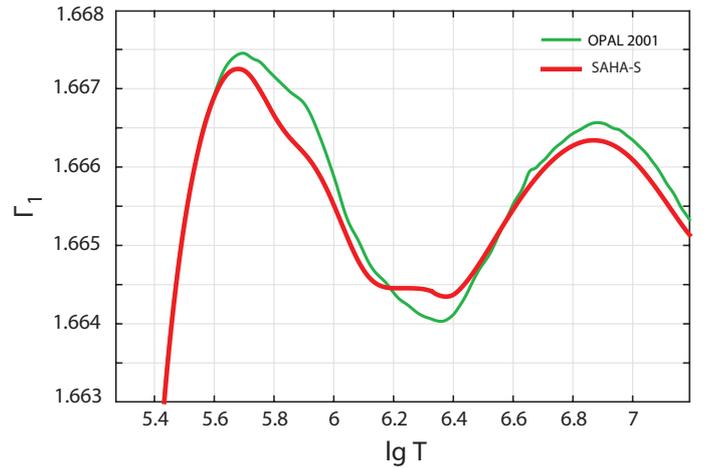}}
        \caption{Adiabatic exponent $\Gamma_1$ in the deeper interior of the Sun (down to the centre)}
        \label{FigGamma1}%
\end{figure}

The adiabatic exponent $\Gamma_1$ of the different models of the present-day Sun is plotted for the deep part of the solar interior (Fig.~\ref{FigGamma1}). The outermost part closer to the surface, which would exhibit large variations in $\Gamma_1$, is not shown. In the deep interior of the convection zone, the difference in $\Gamma_1$ due to EOS can attain an absolute value up to $10^{-3}$. This discrepancy is detectable by present-day helioseismology \citep{Vorontsov2013}.
Generally speaking, the discrepancy in $\Gamma_1$  in Fig.~\ref{FigGamma1} is the result of a difference in the structure of the models. Specifically, a first part of the discrepancy is caused directly by differences in the physical assumptions of the EOS and by a different $Z$ mixture. However, a further change is induced indirectly by those model quantities that depend on the structure variables. The net discrepancy is the sum of these two effects.

\begin{figure}
        \centering
        \resizebox{\hsize}{!}{\includegraphics{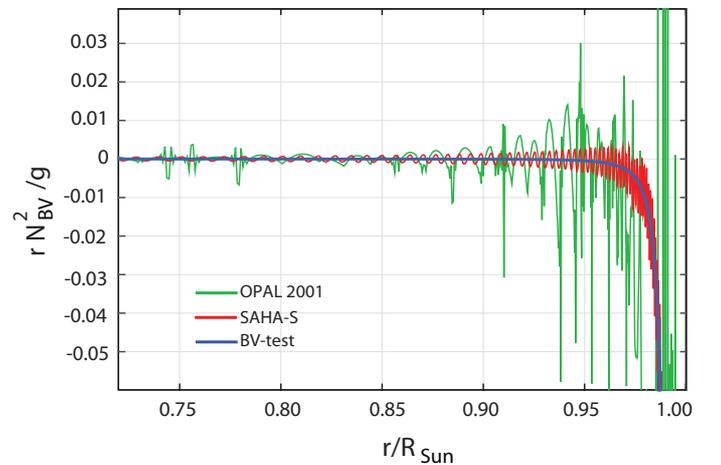}}
        \caption{Dimensionless squared Brunt-V\"ais\"al\"a frequency $rN_\mathrm{BV}^2/g$ in the convection zone of the Sun}
        \label{FigBV}%
\end{figure}

The Brunt-V\"ais\"al\"a frequency $N_{\mathrm{BV}}$ has significant oscillations in the convection zone if it is computed using Eq.~(\ref{EqBVfrequency}), that is, by numerical differentiation of the density. Figure~\ref{FigBV} presents the dimensionless value of $rN_{\mathrm{BV}}^2/g$ computed in models with OPAL 2001 (green curve) and SAHA-S (red curve). The random fluctuations in the case of OPAL 2001 is significantly higher because of higher numerical noise in the OPAL 2001 data tables. The lower level of fluctuations in models with SAHA-S is due to  the quality of our spline interpolation and to less numerical noise in the tables. However, even in the case of SAHA-S, the accuracy of numerical differentiation is not sufficient. In particular, $N_{\mathrm{BV}}^2$ repeatedly changes its sign. The situation can be improved by using Eq.~(\ref{EqBVcz}) in the convection zone (blue curve), which gives a monotonic and negative $N_{\mathrm{BV}}^2$, as dictated by the physics of a slightly sub-adiabatic stratification.

\begin{figure}
        \centering
        \resizebox{\hsize}{!}{\includegraphics{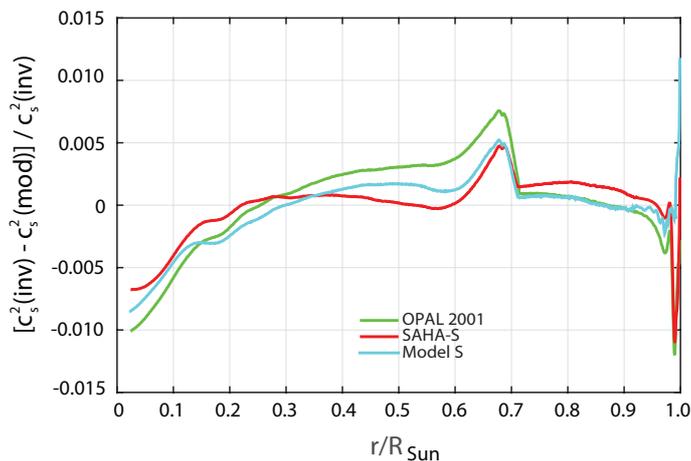}}
        \caption{Relative sound-speed-square difference between a helioseismic inversion \citep{Vorontsov2013} and three different models}
        \label{FigCs}%
\end{figure}

The difference in the sound speed between a model with a given equation of state on the one hand, and the result of a helioseismic inversion \citep{Vorontsov2013} on the other hand, is shown in Fig.~\ref{FigCs}. The red curve refers to a model obtained with the SAHA-S equation of state, the green curve to a model with OPAL 2001. The blue curve represents a standard solar model: Model S of \cite{Christensen1996}, in which an early OPAL version was used \citep{Rogers1996}.
This figure shows how sound speed in modern model calculations closely approximates the observational data, and how accurate EOS data should be to enable an adequate analysis. Inside the convection zone, above the location of $0.713r/R_{\mathrm{Sun}}$, the model structure is predominately defined by $\Gamma_1$, and the differences between the curves reflect the thermodynamic differences between the applied EOS. The required accuracy of the sound speed in EOS calculations has to be better than $10^{-4}$ \citep[see][for a detailed analysis]{Vorontsov2013}.

Most of the difference between the model sound speed below the convection zone is connected to opacities and the general structure of the model core. It is rather difficult to demonstrate specific effects of the EOS here. Common to all solar models is a hump in sound speed just below the bottom of the convection zone. This hump  is probably related to the tachocline and overshooting, both of which are poorly treated by current models.
The analysis of these model differences is well beyond the scope of the present study.

\section{Conclusions}
\label{SectConclusions}

We present the results of the incorporation of the new version of the SAHA-S equation of state into an advanced stellar evolution code, CESAM2k. There is a mismatch between the data available in the thermodynamic calculations and the data needed for stellar modelling. These differences are due to the fact that the evolution code calls for quantities as functions of pressure and temperature instead of temperature and density, which are the most common independent variables in thermodynamic computations. The different independent variables require re-writing the set of partial derivatives of the thermodynamic quantities. To avoid time-consuming and noisy finite-difference numerical differentiations during the calculations, throughout we rely on thermodynamic relations and analytic expressions, and  on pre-computed values of the auxiliary interpolation coefficients $\alpha_i$.
A specific feature of the CESAM2k code is the demand for some extra quantities such as specific heat capacities and adiabatic gradients.

Due to the regular node mesh in the SAHA-S equation of state tables, we were able to use an efficient interpolation algorithm in the form of three-dimensional cubic spline functions, allowing us to keep the smoothness of the functions and their first derivatives, and  to provide good estimates of higher partial derivatives, in particular those with respect to chemical composition.
With respect to interpolation in $Z$, we have adopted linear interpolation between adjacent table points.

Our results obtained for the spline coefficients and the necessary interpolation software are available as a freely downloadable FORTRAN module\footnote{\href{http://crydee.sai.msu.ru/SAHA-S_EOS}{http://crydee.sai.msu.ru/SAHA-S\underline{\hphantom{a}}EOS}}.  

We anticipate that our EOS interface routines can be adapted to other stellar evolution codes.

The SAHA-S equation of state has been integrated in the CESAM2k code, and the first results of solar modelling have already been obtained. We have also carried out a preliminary comparison with the results of previous computations which were based on the OPAL equation of state.


\bibliographystyle{aa} 
\bibliography{saha_bibliogr} 

\end{document}